\documentclass[a4paper]{article}
\usepackage{graphicx,hyperref,amsmath,amssymb,fontenc}
\newcommand{\haak}[1]{\!\left(#1\right)}
\newcommand{\rhaak}[1]{\!\left [#1\right]}

\renewcommand{\imath}{\text{i}}

\begin{document}
\author{Saibal Mitra\\
 Email: smitra00@gmail.com
}
\title{Percolation clusters of organics in interstellar ice grains as the incubators of life} 
\date{\today}
\maketitle

\begin{abstract}
Biomolecules can be synthesized in interstellar ice grains subject to UV radiation and cosmic rays. I show that on time scales of $\gtrsim 10^{6}$ years, these processes lead to the formation of large percolation clusters of organic molecules.  Some of these clusters would have ended up on proto-planets where large, loosely bound aggregates of clusters (superclusters) would have formed. The interior regions of such superclusters provided for chemical micro-environments that are filtered versions of the outside environment. 

I argue that models for abiogenesis are more likely to work when considered inside such micro-environments. As the supercluster breaks up, biochemical systems in such micro-environments gradually become subject to a less filtered environment, allowing them to get adapted to the more complex outside environment. A particular system originating from a particular location on some supercluster would have been the first to get adapted to the raw outside environment and survive there, thereby becoming the first microbe.

A collision of a microbe-containing proto-planet with the Moon could have led to fragments veering off back into space, microbes in small fragments would have been able to survive a subsequent impact with the Earth.
\end{abstract}

\section*{keywords}
Origin of life, Astrobiology, Evolution
\maketitle

\section{Introduction}
While the fundamental biochemical mechanisms of living organisms are well understood,  how these mechanisms arose in the early solar system is not known. The problem here is not so much to explain how particular components such as amino acids, nucleobases etc.\ can arise in a prebiotic setting, rather how the machinery of life that's implemented by such components arose, starting from only these simple components. 

One way of attacking this problem is to consider life as it exists today and try to deconstruct it to see if a simpler, more primitive version could have given rise to it.  This has resulted in the RNA world hypothesis \cite{RNA world}. Here one assumes that RNA instead of DNA encoded genetic information and that RNA molecules were involved in catalytic processes that in modern living organisms are implemented by protein-based enzymes. There is then still a significant gap to bridge between simple prebiotic chemistry and the RNA world.

As little progress has been made in the last few decades, we can ask what prevents us from getting to an explanation of how life came into being.  Walker and Davies have recently argued that the problem is more rooted in basic physics or mathematics than in biochemistry \cite{walk, walkh}.  That the problem transcends biochemistry becomes clear if we considering models of artificial chemistry obtained by replacing the rules of chemistry by some arbitrary set of rules that describe the interaction between some set of building blocks. A challenge formulated in \cite{alife} is to demonstrate an artificial chemistry in which the transition to life appears, which is an unsolved problem.

Intuitively the generic problem here is clear: We cannot see how simple building blocks will assemble themselves into complex machines. If we have an imperfect machine we can imagine that it could evolve to become a better machine. But if we start out with a machine that's too dysfunctional, then this will simply break down, and a collection of building blocks can be interpreted as a totally broken down machine. This intuitive argument has been formulated in a more precise way by Eigen  \cite{eigen}. Here one invokes the fidelity of self-maintaining and replicating machines.

In a simplified version of this argument, one considers a machine with $N$ critical parts such that any defect in these critical parts will cause the machine to stop working. Such defects can arise due to external perturbations, and errors in the self-maintenance process and copying process can also lead to defects. For a set of such systems to be able to undergo exponential growth requires that the number of defects per critical part per reproduction time must be smaller than $1/N$. Then because any system that starts out as some random concoction of building blocks can at best be a very low fidelity replicating machine, the number of critical parts cannot be large.
 
Since a large number of parts is needed to implement the machinery needed for self-repair and copying with high fidelity,  the randomly concocted system would first have to increase the number of critical parts, after which it could undergo Darwinian evolution to reconfigure the new parts to improve the fidelity. While the first part of such a process could conceivable happen e.g.\ due to merger processes, the Eigen limit on the fidelity precludes the second part. Just after a merger process, the fidelity will still be the same as what it was before the merger, while $N$ will have doubled, making it likely that the system will fail to reproduce with a growth factor larger than $1$.

Given this fundamental problem, one may consider if life could have formed spontaneously, despite the astronomically low probability of elementary building blocks assembling themselves to form a living cell, one successful abiogenesis trial can be expected in about $10^{5000}$ trials \cite{spont}. A theory known as panspermia  \cite{hw1,hw2,hw3,hw4,hw5}  makes no assumptions about how life began, but explains the emergence of life on Earth by the arrival of pre-existing life in the cosmos. However if the spontaneous appearance of life from basic chemistry (abiogenesis) is to be invoked then the super-astronomical odds against of about $10^{5000}$ can be overcome in a vast Universe almost infinite in size  \cite{cam}.  The theory's main focus is the way life appeared on Earth from its assumed cosmic orgins, while the origins of life from ordinary chemcial processes are left in the dark. 

However, even if we accept that in a large enough, possibly infinite universe, a spontaneous appearance of microbes from simple chemical compounds is possible, there are other objections against such a hypothesis. The fact that co-enzymes contain nucleotides fits in well within the RNA-World hypothesis \cite{coenz}. These nucleotides could just as well be replaced by some other compound while preserving the ability to be able to bind with the enzymes they currently interact with as far as modern biochemistry is concerned, but in the RNA-World the nucleotides would have been needed for base pairing with ribozymes. Overwhelming odds against spontaneous creation thus continue to exist due the existence of biochemical relics. 

With this evidence against the spontaneous origin of life out of simple compounds in hand, we should turn back to tackling the fidelity problem. Let's consider the simplest system that's just able to reproduce with high enough fidelity to have a growth factor larger than $1$. We can then ask how its simpler ancestral system could possibly have worked to get to high enough fidelity to repair itself and reproduce.  Living organisms maintain their internal states out of thermodynamic equilibrium in a highly dynamic way. Then if we have such a system that is the simplest system that can just about work this way, it follows that the ancestral system would have had to work in a more static way. The role of the static parts would be to protect the dynamic degrees of freedom from perturbations from the environment, as well as play a supportive role for the dynamic degrees of freedom that depend on it. For example, if certain enzymes are lacking in the ancesteral system, then certain fixed structures in the static system would have to catalyze the chemical reactions that are catalyzed by these enzymes.  Then such an ancestral system could in turn have evolved from other ancestral systems with even less dynamic degrees of freedom. The original system that gave rise to life could have been a purely static object.  

This static object could then have been a vesicle within which the dynamic parts would appear later. A lot of work has been done on lipid membranes as such vesicles \cite{lipid}, the problem here is that such membranes in the early stages of life would have had to be simpler than the membranes modern cells use, while we need a far more protective environment than the interior parts of modern cells. Another objection against lipid membranes follows from thermodynamics. Living organisms maintain themselves far from thermodynamic equilibrium using their own dynamic degrees of freedom. If the ancestral system of the simplest would also succeed in doing that with less dynamic degrees of freedom, supported by a static structures then such a static structure should itself be an object that's far from equilibrium. The lack of any dynamic processes to maintain the structure far from equilibrium then implies that the object would have to be in a metastable state. So, the sort of vesicle we're looking for should unlike lipid membranes have a rigid molecular structure. Such vesicles would be able to have fixed molecular structures capable of acting as enzymes, as suggested above. 

The rigidity of the vesicle then implies that the vesicle won't take part in reproductive processes. It will act as a micro-environment within which an entire eco-system of reproducing biochemical systems is housed. Lipid membranes can, of course,  still play a role inside such a micro-environment. The question is then how and where such vesicles could have been forged. Clearly, very far from equilibrium processes are needed to yield far from equilibrium, metastable objects.  In the next section we'll see that a cryogenic space environment can provide for the necessary far out-of-equilibrium conditions.

\section{Long term evolution of molecules contained in ice grains irradiated by UV radiation and cosmic rays}
The finding that certain meteorites contain complex organic compounds that must have formed in space \cite{mrch} has led to a lot of research into biochemical processes in space, see e.g.\ early work by Hoyle and Wickramasinghe \cite{hw1,hw2,hw3,hw4}.  As shown in \cite{ciesla}, UV radiation of small ice grains can lead to the synthesis of organic compounds. UV photons incident on ice grains will create ions and radicals that can then later combine due to heating events to form larger organic molecules. This process was considered for ice grains in the solar nebula in \cite{ciesla},  in experimental work it has been found that organic compounds such as amino acids, and nucleobases can be formed due to such processes \cite{amino, rib, nuc}.

Here we'll consider this process for interstellar ice grains in which organic compounds are captured.  In that case, the grains at a temperature of  $\sim 10$ K will experience sporadic heating events due to interactions with cosmic rays \cite{leger, shen, rob, kal}. A detailed calculation of the average time interval between significant CR-induced heating events is given in \cite{kal}.
 
Small micrometer-sized ice grains will be subject to significant UV exposure per molecule given enough time and provided the dust cloud it is sitting in isn't too dense \cite{ciesla}. One can expect $\sim 5\% $ of the contents of an ice grain to be converted to organics after $\sim 10^6$ years of exposure to the cosmic background of $\sim 10^8$ UV photons per second per $\text{cm}^2$. The upper limit is then the available substrate of simple organic compounds, rather than the photon flux or the available irradiation time  \cite{ciesla}. This then does require heating events that in the interstellar case can be generated by CR interactions, in \cite{kal} it is shown that this occurs in dust grains up to a few tenths of a micrometer radius. Such dust grains can experience heating events of tens of K, up to $\sim 100$ K on time scales ranging from years to thousands of years depending also on the size of the ice grain and the thickness of the ice cover. 
  
The long term evolution due to adding new bonds over and over again, can be modeled as a bond percolation problem. In the beginning there will only be small clusters of molecules; the expectation value of the size of clusters measured by the number of monomers will be independent of the number of molecules in the grain. But at some point we cross the so-called critical point for percolation where the largest cluster will have a size that has an expectation value given by $\sim N^{0.841}$ where $N$ is the total number of monomers in the grain \cite{percm,percy,percyer}. The $\sim 0.1$ micrometer size of the ice grains limits $N$ to $\sim 10^8$. At the percolation threshold the largest cluster can then contain about $5\%$ of the monomers. 

On the simple cubic lattice the critical percolation threshold is at a bond probability of $p_c = 0.2488$ \cite{percm,percy,percyer}, for other percolation models on different 3 dimensional lattices this critical probability is different, but certain exponents such as the one in the power law for the expectation value of the largest cluster, are universal.  Because the percolation threshold is typically much smaller than $0.5$, the cluster has a very open structure; molecules that are not too large can easily move through the cluster. As more and more bonds appear, the cluster grows and becomes denser.  Up to a critical probability $q_c$ it is still possible to traverse the cluster from one side to the other side through a network of pores. As the critical probability $q_c$ is approached the pores become narrower, some pores become blocked; the large network running through the cluster becomes sparser. Above the critical probability $q_c$ the network spanning the cluster will have collapsed into a large number of smaller isolated networks.
 
 The fact that below the critical bond probability $q_c$, the monomers can penetrate the cluster and reach most of the boundary, including sites that are located deep inside the cluster, means that the cluster will have a homogeneous composition. Above the bond probability of $q_c$, there will be monomers trapped inside isolated network of pores, henceforth referred to as "chambers". There may also be smaller clusters inside chambers. When a CR hits the grain the cluster is located in, the heating of the cluster can make boundaries between chambers permeable to smaller molecules. So, on long time scales, the monomers will be able to traverse the cluster by moving from chamber to chamber. Monomers arriving in one chamber this way will typically spend a long time there before the next CR-induced heating event happens. During that time, the monomers will be subject to chemical reactions.

We thus see that despite the transport of monomers through the network of pores, the composition of the monomers and other chemical compounds in the chambers will depend on the depth inside the cluster, where the depth of a chamber is defined by the minimum number of CR-induced path crossings needed to reach the chamber from the cluster's exterior. This means that the cluster will start to become inhomogeneous, the deeper chambers will close up less fast than the outer chambers.

So, what we've ended up with, is a cluster containing a large number of chambers that are mutually out of chemical equilibrium with each other. It is also out of equilibrium in a stable way, the chemical potentials of the various monomers in the chambers are the result of a long term dynamical equilibrium situation. The chemical potential differences drive chemical reactions when there is diffusion through the chamber boundaries induced by CR interactions. In the long period between CR interactions the chemical potential differences get restored to the old values due to the different ways the chemistry works in the different chambers when they are completely isolated from each other. 

\section{Evolution of clusters in the solar system}
Interstellar ice grains would have ended up in the molecular cloud that gave rise to the solar nebula. Ultimately they would have ended up becoming part of comets and proto-planets in the outer solar system via accretion, while in the inner solar system such ice grains would have evaporated.

Clusters formed inside such grains would have been able to survive intact in a wide range of environments on icy bodies in the outer solar system, but the survival period would be limited by the thermal and chemical properties of the new environment. E.g.\ in water at room temperature, peptide bonds will last for a few centuries \cite{pep}.  A cluster ending up inside a proto-planet would no longer be subject to UV radiation and it would be shielded from CRs in the MeV range of energies. It would still be subject to secondary CRs such as muons generated by a much higher energy primary CR, but with the cluster in thermal contact with a much larger body, there is then no longer a large CR-induced heating effect.
 
Instead, asteroid impacts would cause localized heating events. Clusters located at an impact site would be destroyed, while farther away the milder heating effects would only cause the ice grains to melt. Clusters liberated from ice grains would then be able to flow away until the temperature of the local environment dropped back below freezing point.  During this process, clusters would occasionally interact with each other, sometimes binding to each other via weak bonds. On the long term, after many impact generated heating events, large aggregates of clusters (superclusters),  much larger than a micrometer in size would have formed. The structure of a supercluster is similar to that of a cluster, except that at the cluster boundaries there are cavities that are much larger than the chambers. Unlike the clusters, such a supercluster is a weakly bound structure that's prone to break-up. 

Chemical compounds from the environment moving into the outer chambers of a supercluster may be able to percolate into inner chambers and inner cavities, depending on the nature of the chemical compounds and the temperature. As the chemicals pass from chamber to chamber into the very deep chambers and cavities, they get filtered and also modified due to chemical reactions. The chemical potentials of the monomers deep inside the supercluster will therefore be determined by the long term average of the exposure of the supercluster to its environment; rapid fluctuations of the chemicals it is exposed to, will be damped down.

Superclusters would have been exposed to cold-warm cycles due to the rotation of the proto-planet around its axis, the amplitude of such cold-warm cycles would be modulated due to the eccentricity of the orbit.  The migration of Jupiter due to its interaction with the solar nebula  (the so-called "Grand Tack") \cite{tack}  led to a fraction of the proto-planets from the outer solar system to be scattered into the inner solar system \cite{water}.  Such proto-planets would have ended up in orbits with large eccentricities, the superclusters in such proto-planets would therefore have been subject to cold-warm cycles that became much stronger than average near the perihelion of the orbit. 

Deep inside these superclusters, the chemical potentials could thus have been a purely periodic function of time with a period of one proto-planet year. At lesser depths, the signal due to the shorter term cold-warm cycle caused by the rotation if the proto-planet would dominate. At the outer regions of the supercluster, the chemical potentials could have exhibited strong random behavior on short time scales due to the complex chemical environment the supercluster was in.

\section{Abiogenesis inside superclusters}
Many of the proposed models for abiogenesis that have been proposed in the context of prebiotic Earth, can be considered inside the more protected chemical environment found inside the deep interior of a supercluster. E.g.\ we may consider the scenario of a protocell containing RNA \cite{soz}. This requires RNAs with random information content and fatty acid molecules to form the cell membranes. These molecules might have formed when the cluster was forged in the interstellar environment. In a suitable cold-warm cycle, with enough nucleobases and fatty acid molecules present, the RNAs will be able to copy itself after splitting in the warm phase followed by a template copying process in the cold phase. 

According to the conventional protocell model, the evolution of the RNA is driven by its potential to control the permeability properties of the membrane \cite{soz}. If due to random mutations, RNA appears that is able to modify the membrane such that the flow of molecules in and out of the protocell leads to a faster replication rate, then these protocells will dominate in the next generation. This proposed evolution toward a more functional RNA is more likely to work in a chemical environment that changes in a simple, predictable way. A cavity in the deep interior of a supercluster would have been a more suitable micro-environment for protocells to evolve to control its cell membrane. Another advantage of such a micro-environment is that the chambers that are connected to the cavity would have acted as reaction chambers over which the protocells would have some control over. 

To get to a gradual adaptation to the outside environment, one can consider a scenario where the weakly bound supercluster breaks up. One may e.g.\ consider a gradual change in the orbit of the proto-planet that the supercluster is located in, moving it closer toward the Sun. Gradually rising average temperatures may then cause a gradual break-up of the supercluster, thereby exposing the protocells gradually to a less filtered chemical environment. 

Many other models for abiogenesis can be considered instead of the protocell scenario. In general, the break-up of the supercluster with the proposed primitive biochemical systems staying confined in the same local area in the interior, provides for a smooth path for adaptation to the outside environment. 

From different parts in the same superclusters or from other superclusters, different lifeforms might have evolved. However, the vast majority of such systems will not yield  lifeforms capable of surviving outside of the supercluster. E.g.\ if the supercluster breaks up too fast this may not yield a successful lifeform. A more gradual breakup process where the supercluster gradually sheds clusters from the outer layers is more likely to yield success. The first such lifeform capable of surviving in the outside environment would colonize the local environment and would likely have out-competed other lifeforms that emerged later.

\section{Transfer of microbes to the Earth}
If a simple biochemical  process starting in the interior of a supercluster eventually leads to microbes in the raw environment of a proto-planet, then the next question to address is how such microbes would have ended up on Earth. The earliest evidence for life has been dated to between $3.770$ billion and possibly $4.280$ billion years old \cite{age}. This means that sophisticated microbes existed at least at the end of the so-called Late Heavy Bombardment period (LHB) that lasted from  from 4.1 billion years ago till 3.8 billion years ago \cite{lhbage}. The LHB was caused by an orbital resonance between Saturn and Jupiter after the end of the Grand Tack, causing an influx of asteroids into the inner solar system \cite{lhb}. If, as is suggested in this article, abiogenesis occurred on a proto-planet, then that would suggest that life appeared on Earth during the LHB as a result of an impact by a proto-planet. 

A problem here is that while it has been shown that meteorites can carry microbes to a planet \cite{shp}, the impacts due to collisions with proto-planets are probably too violent to be able to deliver living organisms to the Earth. Such impacts would cause enormous heating of the Earth's surface and atmosphere, all the oceans would likely evaporate \cite{evap}. So, even if any lifeforms made it to Earth alive, they would be unlikely to survive. However, as pointed out in \cite{shultz}, a close investigation of Mare Imbrium on the Moon shows that this impact was caused by an oblique impact by a proto-planet of approximately $250$ km diameter, of which huge fragments broke off and veered back into space. Microbes in such fragments would then survive, and if they are in small enough fragments of the order of several meters diameter or smaller, they would also survive after being delivered to Earth later  \cite{shp}. During the LHB, the Moon was closer to Earth, at about $1.4\times 10^5$ km distance \cite{dist},  impacts on the Moon would have generated a significant flux of lunar rock and impactor fragments incident on the Earth's surface.

The probability $p$ for a random impact by an impactor of radius $r$ to be an oblique impact is easily estimated. The available cross section for oblique impacts is $\pi \rhaak{R^2 - (R-r)^2} =\pi\haak{2 R r - r^2} $ where $R= 3474.2 \text{ km}$ is the lunar radius. Dividing this by the total cross section of $\pi R^2$ available for the impact, yields:
\begin{equation}
p = 2 \frac{r}{R} - \haak{\frac{r}{R}}^2
\end{equation}
This evaluates to approximately $0.07$ for an impactor of diameter $250\text{ km}$, so it's not an extremely unlikely scenario for a few such impacts to have happened on the Moon.

Living organisms transported to Earth during the LHB would have had to endure future giant impacts. The Earth should have been hit by $\sim 16$ objects larger than the largest object to hit the Moon \cite{evap} and an Imbrium-sized impact on the Earth would be $\sim 16,000$ times more energetic than the KT impact. But any lifeforms arriving on Earth would likely have had enough time to colonize the Earth before the Earth would suffer a giant impact. There would then have been plenty of opportunity for some of the organisms to have migrated below the surface where they could have survived the consequences of such giant impacts \cite{surv}.

\section{Results and discussion}
In this article, I've proposed a general framework for abiogenesis.  The key points are:
\begin{itemize}
\item In interstellar space, simple organic compounds captured in small ice grains were subject to UV radiation and occasional heating due to incident CRs. This induced a bond percolation process that led to large clusters of organic molecules on a time scale of $\gtrsim 10^6$ years.

 \item On a proto-planet, such clusters can merge into loosely bound superclusters. The deep interior of such superclusters can provide for chemical micro-environments in which conventional models of abiogenesis driven by cold-warm cycles can be considered. 

\item Rapid fluctuations in the chemical potentials of certain chemical compounds that can penetrate the supercluster, will be damped down. Long term gradual and periodic changes then dominate, allowing any biochemical systems inside the superclusters to more easily evolve toward exploiting the conditions in their micro-environments, compared to a similar system in the outside environment.

\item As the supercluster breaks up, the system experiences more of the shorter term fluctuations that has more of a random character. The system can then evolve to adapt to these fluctuations, when doing so right from the start might not have worked.  

\item On a small fraction of the superclusters these processes led to microbes capable of surviving in the outside environment.

\item Microbes were transferred to Earth via a collision of a microbe-containing proto-planet with the Moon. Fragments containing microbes resulting from the giant impact rained down on the Earth.
\end{itemize}
Many different alternative scenarios are possible, e.g.\ the way microbes have been transferred to Earth could have been different, or one may speculate that superclusters appeared on the Earth when it was cool enough and that microbes evolved here on Earth. The role played by superclusters, however, is crucial. The fidelity problem discussed in the Introduction, can be resolved by the micro-environments where simple, fragile biochemical systems can be more stable and functional than in the outside environment.

Conventional scenarios where one has to imagine a chemical evolution of biomolecules in a macroscopic environment, have difficulties explaining where the information specifying how the fundamental biochemistry of living organisms works, came from. The laws of nature conserve information; an absence of information about the present state in the past state implies that the information arose as a result of spontaneous symmetry breaking. Such a process will lead to the symmetry being broken in all possible ways, the entire set of these realized possibilities then contains the same information as the past state. This means that different types of biomolecules playing a potential role in different and mutually incompatible biochemistries would have been mixed up together. This would then have frustrated the evolution of any particular type of biochemical system. 

In the presented scenario in this article, the relevant symmetry breaking events occur at the creation of the clusters of different shapes and sizes in interstellar space, when later the clusters merge into superclusters  and when superclusters break up. This means that different biochemical systems would be located in different superclusters or in different parts of the same supercluster. The different processes taking place in the locally different micro-environments there could then proceed without mutual interference. 

The proposed framework for abiogenesis in this article suggests solutions to some of the outstanding problems in artificial life. One unsolved problem is to demonstrate an artificial chemistry in which a transition to life occurs \cite{alife}. Based on the ideas presented in this article, a generic way to get to a successful outcome is to start from an initial state that proved for randomness in the way building blocks form structures. This way one can get to analogues of the superclusters described in this article. However, most artificial life models are based on 2 dimensional cellular automata (CA) where such structures cannot form. In two dimensions, more than one opening in an enclosure will make the structure fall apart. However, it is still possible to use 2 dimensional CAs to simulate systems in their micro-environments. This then requires time dependent ambient conditions of the micro-environment. If successful this would be an interesting result, but this would violate the rules for a successful demonstration as stipulated in \cite{alife}. Therefore, to make progress, simulations must involve 3 dimensional models.

Efforts to create life in the lab have been unsuccessful so far, the results of this article imply that such efforts based on current approaches are doomed to fail. One can try to emulate the processes described in this article, but creating percolation clusters of organics in the lab is difficult. In the proposed scenario, clusters were created on a time scale of $\gtrsim 10^6$ years under a flux of $\gtrsim 10^8$ UV photons per second per $\text{cm}^2$.  An equivalent dose delivered in a few years time would deliver an amount of power of the order of a Watt per kg, making it extremely difficult to maintain the very low temperatures required for the process. Abiogenesis in the lab may, however, be possible without using superclusters. Since in the proposed scenario, the evolving biochemical system keeps on residing at the same area inside the supercluster, one only needs to create a connected system of chambers feeding into a cavity with small openings, allowing small molecules to enter and leave. Such large molecules can be put in a reaction chamber which emulates the chemical conditions just outside the area in the supercluster that is represented by the large molecules. 

The crucial role played by the radiation environment in space has implications for the search for extraterrestrial life. Both UV radiation and CRs are needed to forge the percolation clusters. The CRs at the MeV energies that are needed are mainly produced by supernova \cite{far}, a chain reaction of supernovae where one triggers the collapse of a molecular cloud, leading to the formation of giant stars that then explode as supernovae a short time later \cite{far}, may have led to a higher CR flux around the time of the formation of the solar system. This then suggests that extraterrestrial life is more likely to exists in regions that have undergone a similar process.

\section{Acknowledgment}
I thank Glennys Farrar for pointing out the source of the CRs needed in the proposed scenario.

\end{document}